\documentclass[conference, 10pt]{IEEEtran}

\usepackage[style=ieee, backend=biber, isbn=false]{biblatex}

\newif\ifshowinfo
\showinfotrue

\usepackage[cmex10]{amsmath}
\usepackage{amssymb}
\usepackage{url}
\usepackage{graphicx}
\usepackage{mathtools}
\usepackage{comment}
\usepackage{color}
\usepackage{placeins}
\usepackage{float}
\usepackage{hyperref}
\usepackage{tabularx,colortbl}
\usepackage{ifthen}
\usepackage{cleveref}
\usepackage{siunitx}

\makeatletter

\newcounter{authcount}
\renewcommand{\author}[2][]{
   \stepcounter{authcount}
   \@namedef{author@\theauthcount}{#2}
   \@namedef{authorlabel@\theauthcount}{#1}
}

\newcounter{addrcount}
\newcommand{\address}[2][]{
   \stepcounter{addrcount}
   \@namedef{address@\theaddrcount}{#2}
   \@namedef{addresslabel@\theaddrcount}{#1}
}

\newcommand{\alsep}{and}

\def\newmaketitle{\par%
  \begingroup%
  \normalfont%
  \def\thefootnote{}
  \def\footnotemark{}
  \let\@makefnmark\relax
  \footnotesize
  \footnotesep 0.7\baselineskip
  \normalsize%
  \twocolumn[\thenewmaketitle\@IEEEaftertitletext]%
  \if@IEEEusingpubid
     \enlargethispage{-\@IEEEpubidpullup}%
  \fi
  \endgroup
  \setcounter{footnote}{0}\let\maketitle\relax\let\@maketitle\relax
  \gdef\@thanks{}%
  \let\thanks\relax}

\def\thenewmaketitle{
  \newpage
  \begin{center}%
    \vskip0.2em{\Huge\@IEEEcompsoconly{\sffamily}\@IEEEcompsocconfonly{\normalfont\normalsize\vskip 2\@IEEEnormalsizeunitybaselineskip
   \bfseries\large}\@title\par}\vskip1.0em\par%
    \vspace{1ex}
    \newcounter{authloop}
    \newcounter{c@tmp}
    \ifthenelse{\value{authcount}=2}{%
      \newcommand{\liand}{ and }}{%
      \newcommand{\liand}{, and }}
    \ifthenelse{\value{addrcount}<2}{%
      \@nameuse{author@1}%
      \stepcounter{authloop}%
      \whiledo{\value{authloop}<\value{authcount}}{%
        \setcounter{c@tmp}{\value{authcount}}%
        \addtocounter{c@tmp}{-\value{authloop}}%
        \ifthenelse{\value{c@tmp}=1}{%
          \renewcommand{\alsep}{\liand}}{\renewcommand{\alsep}{, }}%
        \stepcounter{authloop}\alsep \@nameuse{author@\theauthloop}}\\%
    }
{
  \@nameuse{author@1}${}^{(\ref{org1},\ref{org2})}$%
  \stepcounter{authloop}%
  \whiledo{\value{authloop}<\value{authcount}}{%
    \setcounter{c@tmp}{\value{authcount}}%
    \addtocounter{c@tmp}{-\value{authloop}}%
    \ifthenelse{\value{c@tmp}=1}{%
      \renewcommand{\alsep}{\liand}}{\renewcommand{\alsep}{, }}%
    \stepcounter{authloop}\alsep \@nameuse{author@\theauthloop}%
      ${}^{(\ref{\@nameuse{authorlabel@\theauthloop}})}$%
  }
}
    \vspace{0.2ex}

    \ifthenelse{\value{addrcount}>0}{%
      \ifthenelse{\value{addrcount}=1}{
        {\@nameuse{address@1}}
      }
      {
        \newcounter{c@address}

        \begin{center}
        \whiledo{\value{c@address}<\value{addrcount}}
        {
          \refstepcounter{c@address}
            ${}^{(\thec@address)}$\,%
              \label{\@nameuse{addresslabel@\thec@address}}%
              \@nameuse{address@\thec@address}\\ %
        }
        \end{center}
      } 
    }
    {
      \relax
    }
  \end{center}
}

\makeatother

\title{Spectral Filtering of 3D Integral Operators Using Modified Green’s Functions}

\ifshowinfo
\author[org1,org2]{Alessandro Bellusci}
\author[org1]{Viviana Giunzioni}
\author[org3]{Adrien Merlini}
\author[org1]{Francesco P. Andriulli}

\address[org1]{Department of Electronics and Telecommunications, Politecnico di Torino, Turin, Italy}
\address[org2]{Early Research Honors School, Politecnico di Torino, Turin, Italy}
\address[org3]{Microwave Department, IMT Atlantique, Brest, France}

\fi

\ifshowinfo
\hypersetup{pdftitle={Spectral Filtering of 3D Integral Operators Using Modified Green’s Functions}}
\else
   \hypersetup{
       pdfauthor={},
       pdfcreator={},
       pdfproducer={},
       pdftitle={Spectral Filtering of 3D Integral Operators Using Modified Green’s Functions},
       pdfsubject={}
   }
\fi

\begin{document}

\newmaketitle

\begin{abstract}
Several recent contributions have analyzed and illustrated the effectiveness of operator filtering, both in terms of regularization and compression, when handling dense matrices arising from the discretization of integral operators, e.g. the single-layer operator. Previous works have introduced different filtering strategies, ranging from Laplacian-based filters to analytically derived ones, with the goal of improving the computational efficiency of iterative and direct solvers for integral equations in the two-dimensional space, like the 2D Electric Field Integral Equation (EFIE). In this work, we propose a filtering strategy based on the spectral truncation of the kernels of integral operators associated with the 3D EFIE. The approach relies on an appropriate spectral representation of the Green's function obtained via the spherical Hankel transform, which provides an analytical foundation for the proposed approach. Finally, we provide semi-analytical and numerical evidence of the impact of this filtering technique on the spectral properties of continuous integral operators and of their discretization through boundary elements, both for the static and dynamic cases.
\end{abstract}

\section{Introduction}
The Boundary Element Method (BEM) is a well-established numerical method in computational electromagnetics, notably suited to open-region electromagnetic problems, such as wave scattering and antenna radiation. The Electric Field Integral Equation (EFIE) can be discretized using boundary elements to yield a dense system of equations \cite{jin2015theory}. This leads to high memory and computational costs, motivating the use of both iterative and fast solvers to solve the system. However the solution's precision and number of iterations required are strongly affected by the spectral properties of the integral operators. Operator filtering has been shown to be an effective way to regularize and compress integral operators via the manipulation of specific components of their spectra. While some of the first filtering strategies relied on quasi-Helmholtz filters and enabled the building of fast direct solvers for integral equations, analytical filtering methods based on truncating particular spectral representations of the integral operator kernels have been presented for 2D problems \cite{masciocchi2023operatora}.
In this work, we propose a new filtering strategy for three-dimensional integral operators linked with the EFIE, relying on filtered integral kernels.
We leverage the geometric properties of the Green's function, notably its spherical symmetry, to obtain its spectral representation by applying the Hankel transform, which is a special case of the three-dimensional Fourier Transform (FT). As demonstrated in \cite{vico2016fast}, this transform can be successfully used to regularize the FT of the Green's function by truncating the interaction in the physical space beyond certain boundaries, in order to build fast algorithms for the computation of volume potentials. Instead, we investigate the regularization of the Green's function by truncation in the spectral domain, in order to extend the applications of operator filtering to the three-dimensional space.
In particular, we present a spectral analysis of the obtained filtered single-layer operator, after outlining a semi-analytic procedure for computing the eigenvalues of boundary integral operators with modified kernels. The properties and performance of the proposed approach are further illustrated via numerical results, including the analysis of the spectral behavior of the Calderón preconditioned EFIE operators. 

\section{Notation and Background}
Let $\Gamma \subset \mathbb{R}^3$ be a smooth closed surface modeling a
three-dimensional perfectly electrically conducting (PEC) scatterer, immersed in a
medium with real wavenumber $k$ and impedance $\eta$, and characterized by the
outward unit normal vector $\hat{\mathbf{n}}$.
If $\mathbf{E}^{\mathrm{inc}}$ is the incident electric field and
$\mathbf{j}_s$ is the surface current density induced on $\Gamma$, then
\begin{equation}
\eta \mathcal{T}_k \mathbf{j}_s = - \hat{\mathbf{n}} \times \mathbf{E}^{\mathrm{inc}}\,.
\end{equation}
The electric field integral operator $\mathcal{T}_k$ is defined as
\begin{equation}
(\mathcal{T}_k \mathit{f})(\mathbf{r})
\coloneq -\mathrm{i}\,k\,(\mathcal{T}_{A,k}\mathit{f})(\mathbf{r})
+ \frac{1}{\mathrm{i}\,k}\,(\mathcal{T}_{\Phi,k}\mathit{f})(\mathbf{r})\ \quad \text{for all } \mathbf{r} \in \Gamma
\end{equation}
where
\begin{align}
(\mathcal{T}_{A,k}\mathit{f})(\mathbf{r})
&\coloneq \hat{\mathbf{n}} \times \int_{\Gamma}
g_k(|\mathbf{r}-\mathbf{r}'|)\, \mathit{f}(\mathbf{r}')\, \mathrm{d}\mathbf{r}'\,,\\
(\mathcal{T}_{\Phi,k}\mathit{f})(\mathbf{r})
&\coloneq \hat{\mathbf{n}} \times \nabla \int_{\Gamma}
g_k(|\mathbf{r}-\mathbf{r}'|)\, \nabla' \cdot \mathit{f}(\mathbf{r}')\, \mathrm{d}\mathbf{r}'\,.
\end{align}
The free-space Green's function is defined as
\begin{equation}
g_k(|\boldsymbol{r}-\boldsymbol{r}'|)\ = \frac{e^{\mathrm{i} k |\boldsymbol{r}-\boldsymbol{r}'|}}{4 \pi |\boldsymbol{r}-\boldsymbol{r}'|}
\label{eqn:G}
\end{equation}
which, in static conditions (i.e., $k = 0$), reduces to
\begin{equation}
g_0(|\boldsymbol{r}-\boldsymbol{r}'|)\ = \frac{1}{4 \pi |\boldsymbol{r}-\boldsymbol{r}'|}\,.
\label{eqn:Gstatic}
\end{equation}
The quasi-Helmholtz decomposition of the matrix $\mathbf{T}_k$  resulting from the BEM discretization of $\mathcal{T}_k$ with Rao-Wilton-Glisson (RWG) source and rotated RWG test basis functions, is linked to the single-layer and hypersingular operators (see for example \cite[eqn. 37]{adrian2019refinementfree}), defined as
\begin{align}
\mathcal{S}_k f(\boldsymbol{r}) &\coloneq 
\int_{\Gamma} g_k(|\boldsymbol{r}-\boldsymbol{r}'|)\, 
f(\boldsymbol{r}')\, \mathrm{d}\boldsymbol{r}' 
\\
\mathcal{N}_k f(\boldsymbol{r}) &\coloneq 
\frac{\partial}{\partial n}
\int_{\Gamma} \frac{\partial}{\partial n'}\, 
g_k(|\boldsymbol{r}-\boldsymbol{r}'|)\, 
f(\boldsymbol{r}')\, \mathrm{d}\boldsymbol{r}'
\end{align}
discretized respectively with piecewise constant and piecewise linear basis functions.
In this contribution, we focus on these operators. In particular, given a mesh $K$ of $\Gamma$ composed of $N$ triangles, we define a set of $N$ piecewise constant, or patch, basis functions $\{\phi\}_{i=1}^N$ and a set of $N$ piecewise linear, or dual pyramid, basis functions $\{\psi\}_{i=1}^N$ associated to the cells of the mesh and supported on $K$ and the barycentric dual mesh of $K$, respectively. Following the standard Galerkin approach, we discretize the single-layer operator with the patch basis functions and the hypersingular operator with the dual pyramid basis functions, resulting in matrices $\mathbf{S}_k$ and $\mathbf{N}_k$ with elements 
\begin{equation}
    [\mathbf{S}_k]_{ij} = \langle \phi_i, \mathcal{S}_k \phi_j \rangle\quad \text{and}\quad [\mathbf{N}_k]_{ij} = \langle \psi_i, \mathcal{N}_k\psi_j \rangle
\end{equation}
where $\langle f, g \rangle \coloneq \int_\Gamma f  g \, \mathrm{d}\Gamma$. 
Moreover, we define the mixed Gram matrix $\mathbf{G}$ with elements
\begin{equation}
    [\mathbf{G}]_{ij} = \langle \psi_i, \phi_j \rangle
\end{equation}
useful for the definition of the Calderón preconditioned form $\mathbf{G}^{-T}\mathbf{SG}^{-1}\mathbf{N}$.

\section{Operator Filtering through the Green's Function Spectral Truncation}
In this section, we introduce a three-dimensional filtering strategy based on the truncation of the spectral representation of the Green’s kernel
$g_k(|\boldsymbol{r}-\boldsymbol{r}'|)$. We will follow the procedure:
\begin{itemize}
\item[(i)] define the filtered Green's function $g_k^{\alpha}(|\boldsymbol{r}-\boldsymbol{r}'|)$, where $\alpha$ is a filtering parameter (i.e. a cutoff spatial frequency);
\item[(ii)] define the filtered operators:
\begin{align}
\mathcal{S}_k^\alpha f(\boldsymbol{r}) &\coloneq 
\int_{\Gamma} g_k^\alpha(|\boldsymbol{r}-\boldsymbol{r}'|)\, 
f(\boldsymbol{r}')\, \mathrm{d}\boldsymbol{r}' ,
\label{eqn:Salpha}
\\
\mathcal{N}_k^\alpha f(\boldsymbol{r}) &\coloneq 
\frac{\partial}{\partial n}
\int_{\Gamma} \frac{\partial}{\partial n'}\, 
g_k^\alpha(|\boldsymbol{r}-\boldsymbol{r}'|)\, 
f(\boldsymbol{r}')\, \mathrm{d}\boldsymbol{r}' .
\label{eqn:Nalpha}
\end{align}
\item[(iii)] finally, discretize the filtered operators via BEM to obtain the matrices $\mathbf{S}_k^\alpha$ and $\mathbf{N}_k^\alpha$ with their corresponding spectra. 
\end{itemize}

The filtered Green's function is obtained by mapping $g_k(|\boldsymbol{r}-\boldsymbol{r}'|)$ into the spectral domain via a multidimensional Fourier expansion, and then applying an inverse transformation to a truncated spectral representation of the original function. As rigorously illustrated in \cite{sheppard2014greenfunction}, the 3D transform of the Green's function requires careful treatment. For spherically symmetric functions like the Green's function, the 3D Fourier transform reduces to the Spherically Symmetric Fourier Transform (SSFT), also known as the spherical Hankel transform. Let $f(\boldsymbol{\rho})$ = $f(\rho)$, with $\rho = |\boldsymbol{\rho}|$, be a radially symmetric function. Then its Fourier transform $F(\mathbf{s})$, with $s = |\mathbf{s}|$, is also radially symmetric (i.e., $F(\mathbf{s})=F(s)$). The functions $f(\rho)$ and $F(s)$ satisfy
\begin{align}
F(s) &= 4\pi \int_{0}^{\infty} \frac{\sin(s\rho)}{s\rho}\, f(\rho)\, \rho^{2}\, \mathrm{d}\rho .
\\
f(\rho) &= \frac{1}{2\pi^{2}} \int_{0}^{\infty} \frac{\sin(s\rho)}{s\rho}\,F(s)\, s^{2}\, \mathrm{d}s .
\label{eqn:spectralrepr}
\end{align}

\begin{figure}
\begin{center}
    \includegraphics[width=3.5in]{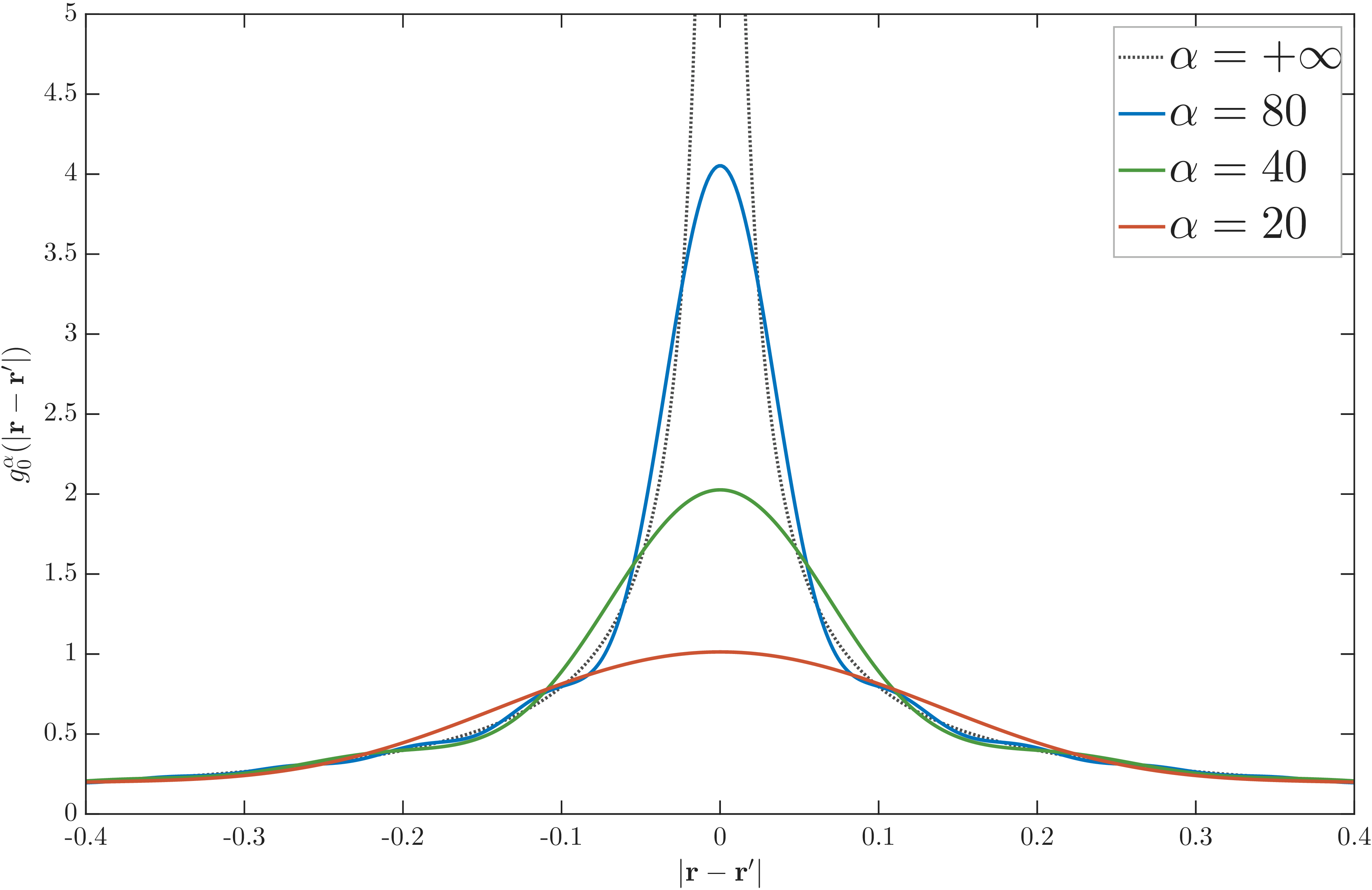}
    \caption{Comparison of $g_0(|\boldsymbol{r}-\boldsymbol{r}'|)$ and $g_0^{\alpha}(|\boldsymbol{r}-\boldsymbol{r}'|)$ for different values of $\alpha$, when $k$ = 0.}
    \label{fig:Gfiltstatic}
\end{center}
\end{figure}
\begin{figure}
\begin{center}
    \includegraphics[width =3.5in]{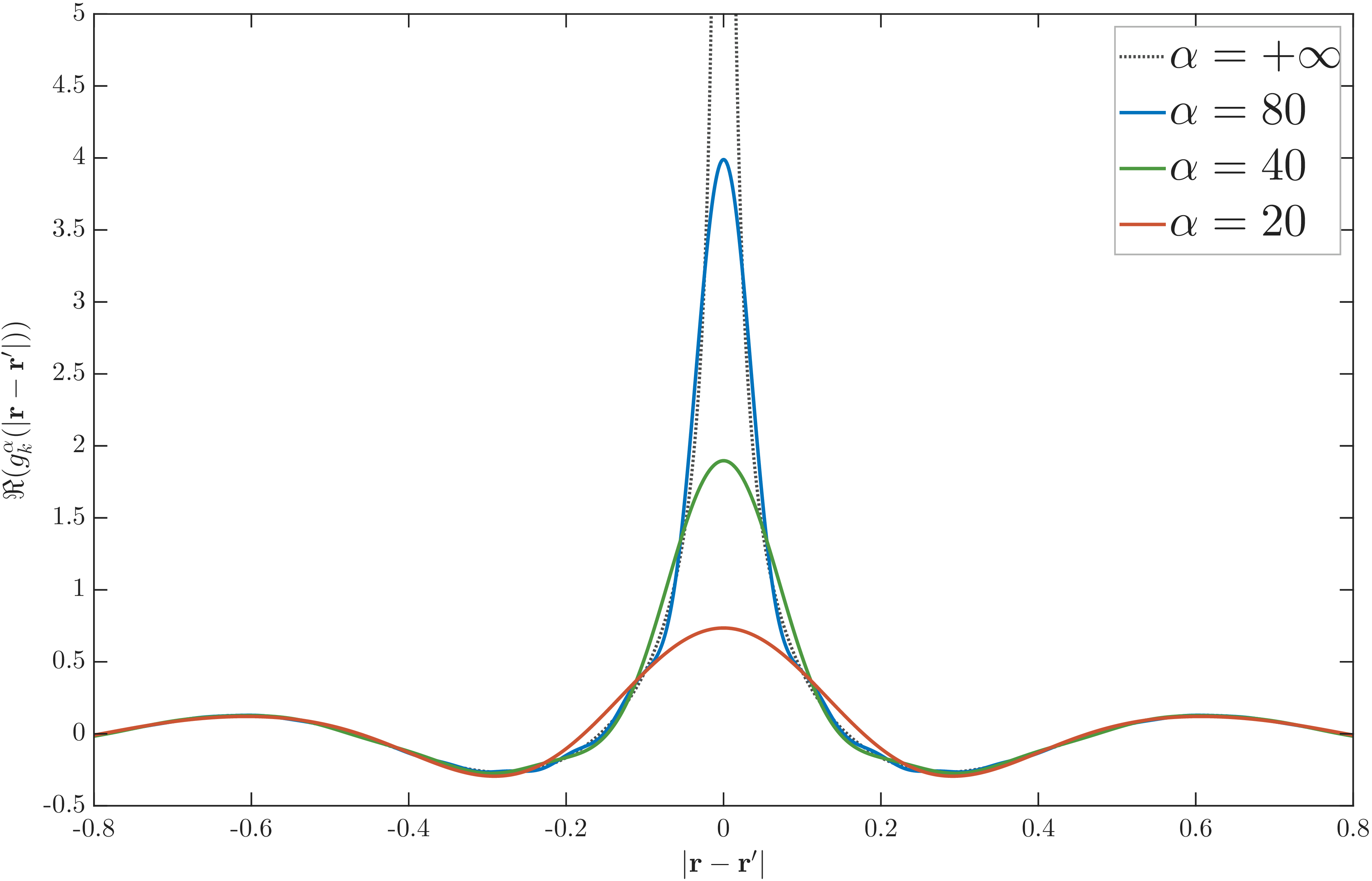}
    \caption{Comparison of $\Re(g_k(|\boldsymbol{r}-\boldsymbol{r}'|))$ and $\Re(g_k^{\alpha}(|\boldsymbol{r}-\boldsymbol{r}'|))$ for different values of $\alpha$, when $k$ = 10 $\text{m}^{-1}$.}
    \label{fig:Gfiltdynamic}
\end{center}
\end{figure}

In the static case, by substituting the standard distributional Fourier Transform of \eqref{eqn:Gstatic} as reported in \cite{sheppard2014greenfunction},
\begin{equation}
G_0(s) = \frac{1}{s^2}\,,
\end{equation}
in \eqref{eqn:spectralrepr}, we obtain a spectral representation of the Green's function kernel
\begin{equation}
g_0(|\boldsymbol{r}-\boldsymbol{r}'|) = \frac{1}{2\pi^{2}} \int_{0}^{\infty} \frac{\sin(s|\boldsymbol{r}-\boldsymbol{r}'|)}{s|\boldsymbol{r}-\boldsymbol{r}'|}\, \mathrm{d}s\,.
\label{eqn:defGstatspectr}
\end{equation}
By truncating the integral in \eqref{eqn:defGstatspectr}, we obtain the filtered Green's function
\begin{align}
g_0^{\alpha}(|\boldsymbol{r}-\boldsymbol{r}'|) &= \frac{1}{2\pi^{2}} \int_{0}^{\alpha} \frac{\sin(s|\boldsymbol{r}-\boldsymbol{r}'|)}{s|\boldsymbol{r}-\boldsymbol{r}'|}\, \mathrm{d}s \\
&= \frac{1}{2\pi^{2} |\boldsymbol{r}-\boldsymbol{r}'|}
\int_{0}^{\alpha |\boldsymbol{r}-\boldsymbol{r}'|} \frac{\sin t}{t}\, \mathrm{d}t\,,
\label{eqn:defGalphastat}
\end{align}
where we have introduced the change of variable which maps $s|\boldsymbol{r}-\boldsymbol{r}'| \to t$. Finally, we can substitute the definition of the sine integral function \cite[eqn. 5.2.1]{abramowitz1964handbook}
\begin{equation}
\mathrm{Si}(x) = \int_{0}^{x} \frac{\sin(t)}{t} \mathrm{d}t
\end{equation}
in \eqref{eqn:defGalphastat} to obtain
 \begin{equation}
g_0^{\alpha}(|\boldsymbol{r}-\boldsymbol{r}'|) = \frac{1}{2\pi^{2} |\boldsymbol{r}-\boldsymbol{r}'|}\,\mathrm{Si}(\alpha |\boldsymbol{r}-\boldsymbol{r}'|)\,.
\label{eqn:Galphasi}
\end{equation}
Using $\lim_{x \to \infty} \mathrm{Si}(x) = \pi / 2$ \cite[eqn. 5.2.25]{abramowitz1964handbook}, we can verify that, for $\alpha \to +\infty$, \eqref{eqn:Galphasi} approaches \eqref{eqn:Gstatic}. 
\Cref{fig:Gfiltstatic} compares the unfiltered Green's function with the filtered Green's functions for different values of $\alpha$.
\begin{figure}
\begin{center}
    \includegraphics[width =3.5in]{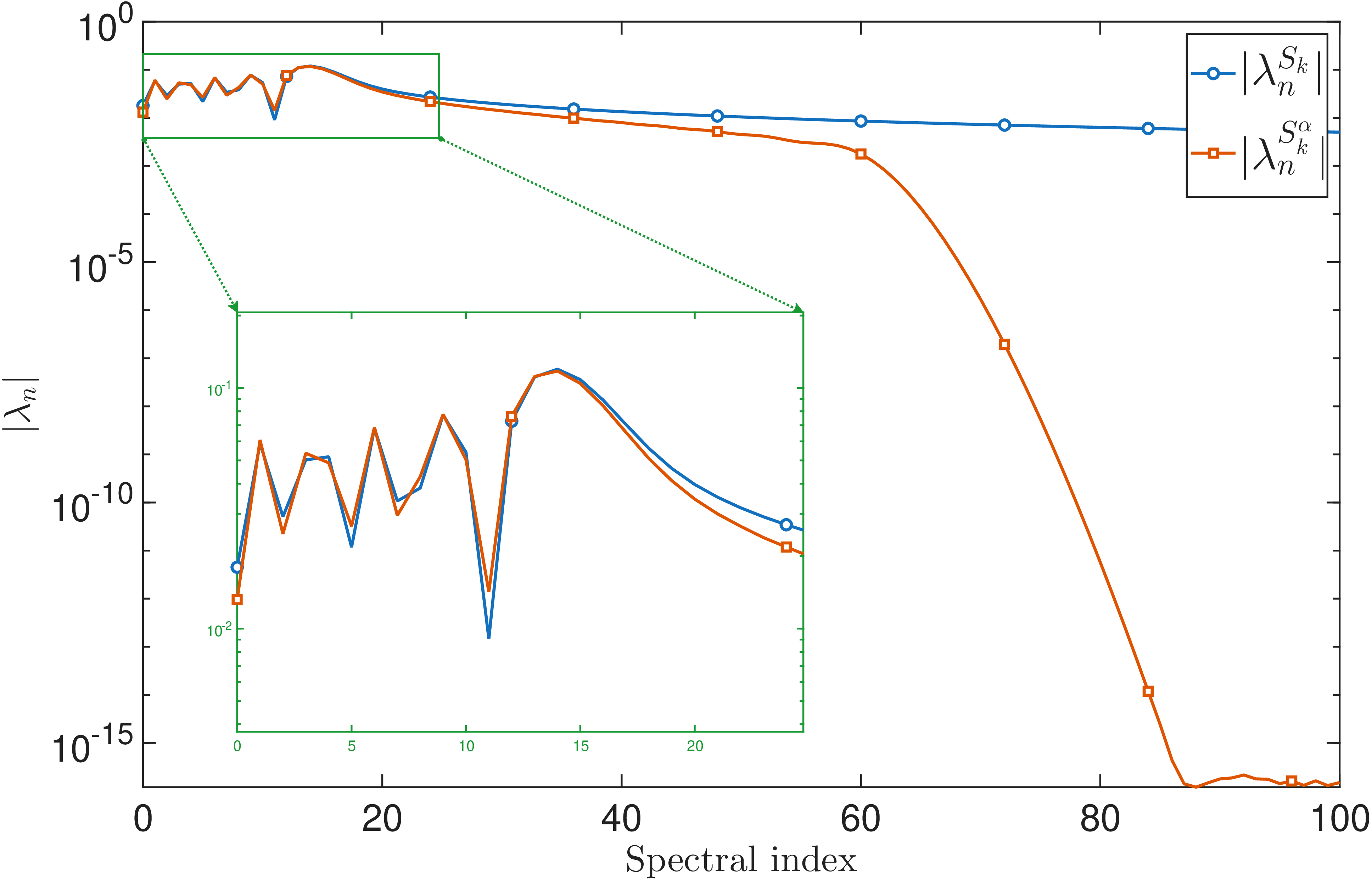}
    \caption{Comparison of the modules of the eigenvalues of operators $\mathcal{S}_k$, computed with \cref{eqn:nonfilteredeigenvalues} and $\mathcal{S}_k^{\alpha}$, computed with \cref{eqn:eigenvaluesfiltered}.}
    \label{fig:analspectrm}
\end{center}
\end{figure}

\begin{figure}
    \centering
    \includegraphics[width=3.5in]{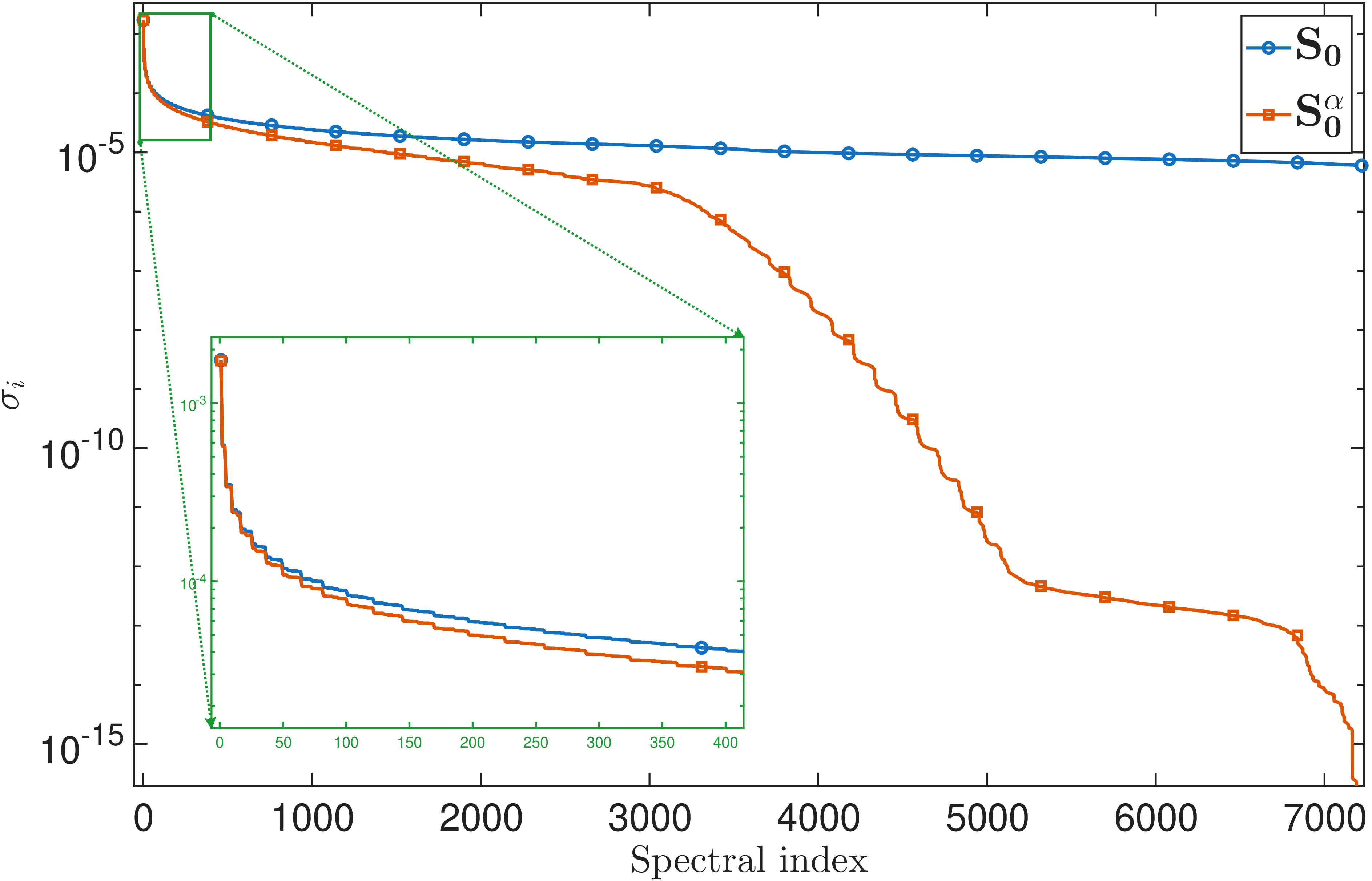}
    \caption{Comparison of the singular values $\sigma_i$ of the matrices $\textbf{S}_0$ and $\textbf{S}_0^{\alpha}$. }
    \label{fig:spectrumSstat}
\end{figure}

A more careful approach is required in the dynamic case to find a spectral representation of the Green's kernel (\cref{eqn:G}).
One possible approach consists in transforming separately its real and its imaginary components; alternatively, the application of the Sokhotsky-Plemelj theorem, as explained in \cite{sheppard2014greenfunction}, directly leads to
\begin{equation}
G_k(s)\ = \mathrm{p.v.} \frac{1}{s^2-k^2} + \mathrm{i} \pi \delta(s^2 - k^2)
\label{eqn:fourierdynamic}
\end{equation}
Physically, the imaginary part consists of propagating waves only and resides on a spherical shell of radius $k$, while the spectrum of the real part, which must be read in the sense of Cauchy principal value, is made up of both propagating and evanescent contributions, and has a singularity in $s = k$.

Following the same approach as for the static case, substituting \eqref{eqn:fourierdynamic} in \eqref{eqn:spectralrepr}, we obtain a spectral representation of the dynamic Green's function. After truncation, this leads to the definition of the filtered kernel $g_k^\alpha$, which can then be evaluated numerically.
As in the static case, also in the dynamic case $g_k^{\alpha}(|\boldsymbol{\rho}-\boldsymbol{\rho}'|)$ approaches  $g_k(|\boldsymbol{\rho}-\boldsymbol{\rho}'|)$ when $\alpha \to +\infty$, as shown in \Cref{fig:Gfiltdynamic}.

Once $g_k^{\alpha}$ has been obtained, we can define the filtered operators $\mathcal{S}_k^{\alpha}$ and $\mathcal{N}_k^{\alpha}$ as in \eqref{eqn:Salpha} and \eqref{eqn:Nalpha}. Their spectral properties differ from those of $\mathcal{S}_k$ and $\mathcal{N}_k$, as will be analyzed in \Cref{sec:spectralanalysis}. Finally, these operators can be discretized using boundary elements, resulting in the matrices $\mathbf{S}_k^\alpha$ and $\mathbf{N}_k^\alpha$.

\section{Spectral Analysis and Numerical Results}
\label{sec:spectralanalysis}
To illustrate the effectiveness of the proposed formulations, semi-analytical and numerical results for both the static and dynamic cases are presented below.

First, we analyze the spectrum of the filtered single-layer operator over the unit sphere $S$ before discretization, to allow for a comparison with its unfiltered version. To compute the eigenvalues $\lambda_n^{\mathcal{S}_k}$ of $\mathcal{S}_k$ and $\mathcal{S}_k^{\alpha}$, we need to expand their kernels in spherical harmonics. 
In the standard single-layer operator case, this leads to the well-known results \cite{hsiao1994error}
\begin{equation}
    \lambda_n^{\mathcal{S}_k} = \begin{cases}
        \frac{1}{2n+1}\,\quad &\text{if $k=0$}\\
        \mathrm{i} k\, j_{n}(k)\, h_{n}^{(1)}(k)\,\quad &\text{if $k>0$}
    \end{cases}
\label{eqn:nonfilteredeigenvalues}
\end{equation}
where $j_n$ is the spherical Bessel function of order $n$, $h_n^{(1)}$ is the spherical Hankel function of the first-type and order $n$ \cite{abramowitz1964handbook}.

When handling boundary integral operators involving the non-standard kernel $g_k^\alpha$, we can leverage the Funk-Hecke theorem which, in the 3D case, states that, if $f$ is a continuous function on $[-1,1]$,
$\boldsymbol{r}$ and $\boldsymbol{r}'$ are two points on the unit sphere $S$, and $Y_n$ is a spherical harmonic of order $n$ \cite{abramowitz1964handbook}, then \cite{han2012integral}
\begin{equation}
\int_{S} f(\boldsymbol{r} \cdot \boldsymbol{r}')\, Y_n(\boldsymbol{r}')\, \mathrm{d}S(\boldsymbol{r}')
= \lambda_n\, Y_n(\boldsymbol{r}).
\end{equation}
with eigenvalue
\begin{equation}
\lambda_n = 2 \pi \int_{-1}^{1} P_{n}(t)\, f(t)\ \, \mathrm{d}t\,,
\end{equation}
where $P_n(t)$ is the Legendre polynomial of degree $n$ \cite{abramowitz1964handbook}. Hence, rewriting $|\boldsymbol{r}-\boldsymbol{r}'|$ as $\sqrt{2 -2(\boldsymbol{r} \cdot \boldsymbol{r}')}$, we can define the eigenvalues of $\mathcal{S}_k^{\alpha}$ as 
\begin{equation}
\lambda_n^{\mathcal{S}_k^\alpha} = 2 \pi \int_{-1}^{1} P_{n}(t)\, g_k^{\alpha}(\sqrt{2 -2t})\ \, \mathrm{d}t\,,
\label{eqn:eigenvaluesfiltered}
\end{equation}
with $t = \boldsymbol{r} \cdot \boldsymbol{r}'$.
The eigenvalues $\lambda_n^{\mathcal{S}_k}$ and $\lambda_n^{\mathcal{S}_k^\alpha}$ of the standard and filtered single-layer operator over the unit sphere $S$ are represented and compared in \Cref{fig:analspectrm}. To obtain these results, the wavenumber has been set to $k = \SI{16}{\per\meter}$ and the spectral cutoff index to $\alpha = 4k$.

\begin{figure}
    \centering
    \includegraphics[width=3.5in,trim=25 1mm 25mm 0,clip]{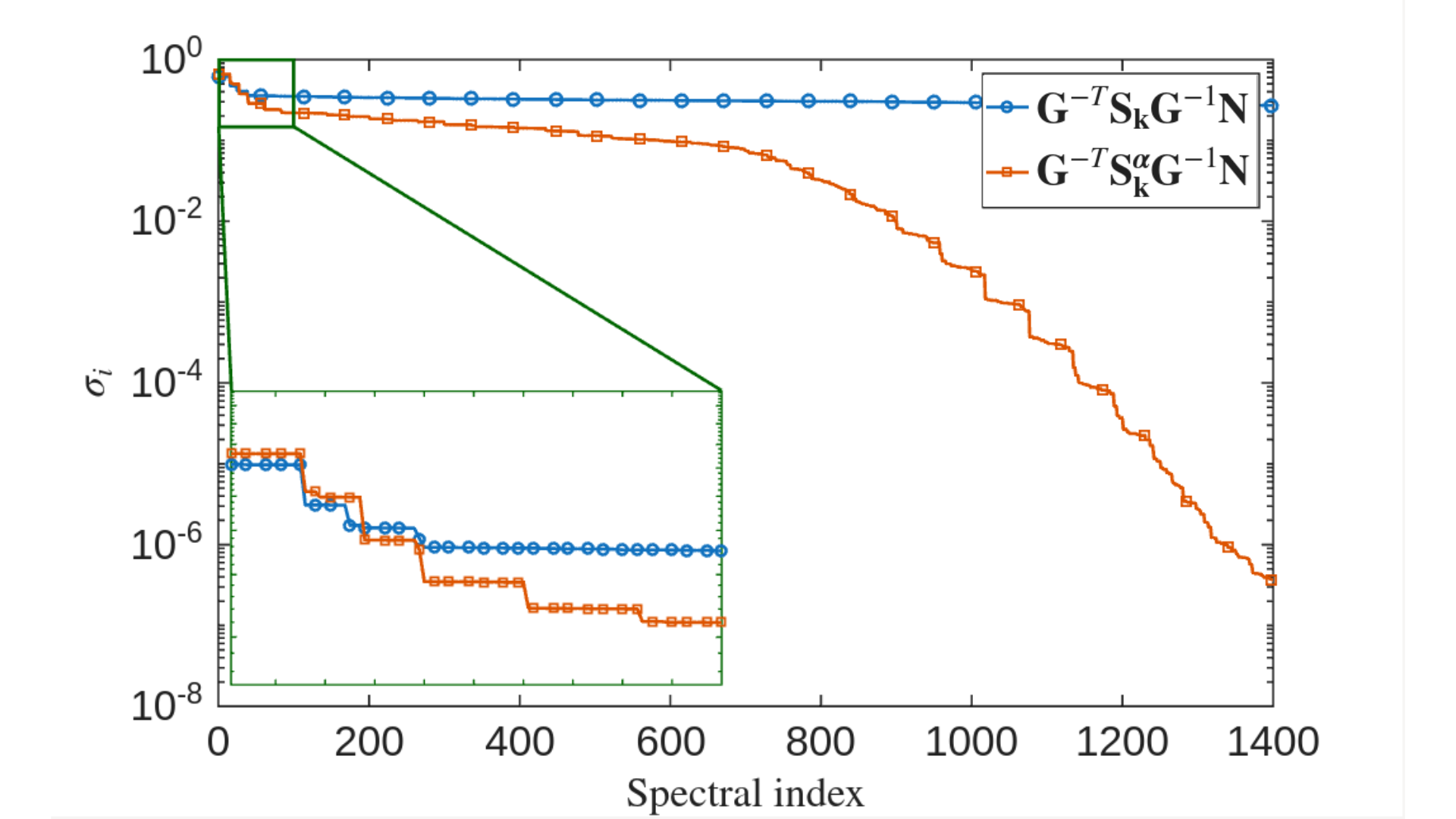}
    \caption{Comparison of the singular values of $\mathbf{G}^{-T}\mathbf{S}_k\mathbf{G}^{-1}\mathbf{N}$ and of $\mathbf{G}^{-T}\mathbf{S}_k^{\alpha}\mathbf{G}^{-1}\mathbf{N}$.}
    \label{fig:spectrumCald}
\end{figure}

Secondarily, we want to analyze the effects of the discretization over the spectra of the operators. To this end, in \Cref{fig:spectrumSstat} the singular values of the single-layer operator matrices $\mathbf{S}_k$ and $\mathbf{S}_k^\alpha$ defined over the triangulation of $S$ are reported, when setting  $k = 0$ and the average edge size of the mesh $h = \SI{0.063}{\meter}$.\\
Finally, we show the effectiveness of the proposed filtering procedure within a Calderón-preconditioned EFIE framework.
In \Cref{fig:spectrumCald}, in particular, we show the singular values of the matrices $\mathbf{G}^{-T}\mathbf{S}_k\mathbf{G}^{-1}\mathbf{N}$ and $\mathbf{G}^{-T}\mathbf{S}_k^{\alpha}\mathbf{G}^{-1}\mathbf{N}$, sorted in decreasing order, with $k = \SI{10}{\per\meter}$, $\alpha = 6k$, and a mesh size parameter $h \simeq \SI{0.13}{\meter}$. These matrices represent discretizations of the Calderón-preconditioned scalar form associated with the quasi-Helmholtz decomposition of the electric field integral operator \cite{adrian2019refinementfree}.
These results hence support the applicability of our filtering procedure even in a Calderón preconditioning framework.

\section*{ACKNOWLEDGEMENT}
\ifshowinfo
This work has received funding from the European Innovation Council (EIC) through the European Union’s Horizon Europe research Programme under Grant 101046748 (Project CEREBRO).
\fi
{ \printbibliography }



%

\end{document}